## Original Paper

**Authors:**
Vishal Dey, BS[1]
Peter Krasniak, MD[2]
Minh Nguyen, MD[2]
Clara Lee, MD[2]
Xia Ning, PhD[1,3, 4]

**Affiliations:**
1. Computer Science and Engineering, The Ohio State University, Columbus, OH
2. Plastic and Reconstructive Surgery, The Ohio State University, Columbus, OH
3. Biomedical Informatics, The Ohio State University, Columbus, OH
4. Translational Data Analytics Institute, The Ohio State University, Columbus, OH

*Corresponding Author
Xia Ning
1800 Cannon Drive, 310C, Columbus, OH 43210
ning.104@osu.edu
614-366-2298

# A Pipeline to Understand Emerging Illness via Social Media Data Analysis: A Case Study on Breast Implant Illness

## Abstract

**Background:** A new illness could first come to the public attention over social media before it is medically defined, formally documented or systematically studied. One example is a phenomenon known as breast implant illness (BII) that has been extensively discussed on social media, though vaguely defined in medical literature.

**Objectives:** The objective of this study is to construct a data analysis pipeline to understand emerging illness using social media data, and to apply the pipeline to understand key attributes of BII.

**Methods:** We conducted a pipeline of social media data analysis using Natural Language Processing (NLP) and topic modeling. We extracted mentions related to signs/symptoms, diseases/disorders and medical procedures using the Clinical Text Analysis and Knowledge Extraction System (cTAKES) from social media data. We mapped the mentions to standard medical concepts. We summarized mapped concepts to topics using Latent Dirichlet Allocation (LDA). Finally, we applied this pipeline to understand BII from several BII-dedicated social media sites.

**Results:** Our pipeline identified topics related to toxicity, cancer and mental health issues that are highly associated with BII. Our pipeline also shows that cancers, autoimmune disorders and mental health problems are emerging concerns associated with breast implants based on social media discussions. The pipeline also identified mentions such as rupture, infection, pain and fatigue as common self-reported issues among the public, as well as toxicity from silicone implants.

**Conclusions:** Our study could inspire future work studying the suggested symptoms and factors of BII. Our study provides the first analysis and derived knowledge of BII from social media using NLP techniques, and demonstrates the potential of using social media information to better understand similar emerging illnesses.

## Introduction

The ubiquity of social media has resulted in the early descriptions of new and evolving diseases over social media platforms before they can be systematically studied [1–7], particularly during the era of medical internet.[8–14] Social media users increasingly turn to platforms like Twitter, Facebook, YouTube, etc., to either share personal experiences, including diseases and illness they experienced, or to seek support and resources, including health and medical resources. Recent studies showed the potential of social media in detection of mental illness and depression [15–17], in early detection of food-borne illnesses [18–20] and other infectious diseases.[2,21–24] Furthermore, several studies demonstrated social media as an effective tool to disseminate information regarding symptoms, personal well-being and public health resources during multiple influenza outbreaks.[25–28] During the early stages of COVID-19, studies [4,29,30] analyzed Sina Weibo (a major Chinese microblogging site) posts to characterize patient symptoms and public concerns in multiple provinces of China. Based on the analysis from Weibo posts, Huang et al.[30] concluded that most of the affected patients were elderly with fever as the most common symptom. These studies demonstrate that public social media data can be leveraged to better understand emerging illnesses and to accommodate prompt responses.

One particular new illness we study in this manuscript is Breast Implant Illness (BII). Breast implants have gained popularity over the last 20 years.[31] More than 400,000 women have breast augmentation or post mastectomy surgeries every year in the US.[32] There has been a 4% increase in the number of breast augmentation procedures during 2017-2018, and over the same period a 6% increase in breast implant removal procedures.[32] Concerns about the safety of breast implants have also arisen [33–38] and persisted.[39–45] However, a causal link between breast implants and systemic diseases has not been definitively shown, yet a phenomenon called “breast implant illness”, which attributes systemic symptoms to breast implants, has emerged.[46] Unlike other new medical illnesses, however, BII has been reported minimally in the medical literature, but has primarily come to attention on social media.[11,47–50] For example, a recent analysis [49] demonstrated increasing public interest in BII based on Twitter and Google Trends data from February of 2018 to 2019. In an attempt to summarize key symptoms, diseases and disorders defining BII, several cohort studies [51,52] analyzed patient reported outcomes before and after breast explant surgeries. These studies showed some potential relations between explant surgeries and improvement of specific symptoms in the patient population. Unfortunately, these studies were not definitive due to limited study design secondary to their lack of control groups, data collection bias, and lack of randomization. The lack of medical knowledge about BII makes it difficult to define the condition and therefore nearly impossible to conduct rigorous epidemiological or clinical studies of it. BII is just one disease process for which the lack of medical knowledge is apparent, but there are many other new illnesses for which this is the case; any initial knowledge that is supported by a sufficient amount of social media data would be meaningful as reference for future, formal studies, and thus, techniques to discover such knowledge are highly needed.

Toward identifying and summarizing key attributes of a new illness, in this study, we constructed a data analysis pipeline for social media data analysis about BII. The pipeline incorporates Natural Language Processing (NLP) and topic modeling methods. Our primary contribution is on deriving novel knowledge about BII – a medical condition which has not yet been systematically studied and defined in the medical literature via constructing a data analysis pipeline and applying the pipeline on social media data. Given the fact that the medical knowledge and literature on BII have not been established, and the related concepts are not well defined or accepted, using social media data to understand emerging issues could be a meaningful starting point. We applied this pipeline to better understand the symptoms and signs that are associated with BII. Our study provides the first analysis, to the best of our knowledge, using social media data, and derived knowledge of BII from social media. It demonstrates the potential of using social media information to better understand conditions that have primarily been reported on social media. It also demonstrates the effectiveness of our pipeline and its potential to be applied to understand other new illnesses. In the following discussion, we will discuss our pipeline within the context of BII. However, our pipeline is not specific to BII and is applicable to other illnesses.

# Methods

## Data

We collected and used data from the following social media websites. These websites were selected because they are dedicated for BII discussions and information and have focused user groups with interest in BII. Often, dedicated social media websites (e.g., forums, twitter sites) are available for a particular illness or disease. For example, some dedicated websites [53–55] contain stories and experiences of patients fighting different cancers; there are dedicated websites [56,57] containing various posts and stories by users experiencing chronic pain and illness; others [58–60] contain stories and experiences from COVID-19 survivors. Below are the social media sources used in our study:

- https://www.breastimplantillness.com: this is a dedicated, public website with articles on BII-related topics, and offers resources related to implant and explant procedures, etc. This website also allows individuals to post their experiences and concerns on breast implants and related health issues. We extracted individuals' posts from the website (up to 05/10/2019), and the resulted dataset is referred to as BIIweb.
- https://healingbreastimplantillness.com: this website contains information on post-implant disorders, post-explant healing, breast implant safety, etc. The discussion board of this website has multiple posts and comments on symptoms, signs, etc., that are experienced by individuals with a breast implant or those who have undergone an explant. The dataset extracted from the discussion board of this website (up to 05/10/2019) is referred to as HealingBII.
- https://www.instagram.com/explore/tags/breastimplantillness: this website contains the collection of publicly available Instagram posts that used "breastimplantillness" as a hashtag in their posts. We extracted the associated texts of each Instagram post with the timestamp between 01/10/2012 and 09/04/2019. The dataset extracted from this site is referred to as IG-BII.

Table 1: Statistical Summary of Social Media Data Analyzed

| Dataset | #posts[a] | $l_{max}$[b] | $l_{min}$[c] | $l_{avg}$[d] | #words[e] |
|---|---|---|---|---|---|
| BIIweb | 187 | 669 | 3 | 129 | 24,191 |
| HealingBII | 1,920 | 1,330 | 1 | 85 | 165,090 |
| IG-BII | 28,987 | 515 | 1 | 123 | 3,581,081 |

In this table, [a]#posts indicates the number of posts/comments in the respective datasets.
[b]$l_{min}$/[c]$l_{max}$ indicates the maximum/minimum length of a post in terms of words.
[d]$l_{avg}$ indicates the average length of posts in terms of words.
[e]#words indicates the total number of words in respective datasets.

All comments/posts from the three websites are included in the corresponding datasets. Table 1 presents a summary of the collected social media data. The BIIweb dataset has only 187 posts but larger posts (larger $l_{avg}$) on average than the other two datasets. HealingBII is the second largest dataset with 1,920 posts, each with 85 words on average. IG-BII is the largest dataset with 28,987 posts and 123 words per post on average.

## The Pipeline

Figure 1 presents the overview of the pipeline. We extracted major topics of interest primarily related to symptoms, diseases and medical procedures from our datasets through the following three steps. Each of the steps will be discussed in detail later.

Step 1. Data preprocessing: We removed all stop-words, numeric characters, hyperlinks, hashtags, etc., and converted all the remaining characters into lowercase.

Step 2. Mention extraction and concept mapping: We extracted mentions related to signs/symptoms, diseases/disorders and medical procedures using the Clinical Text Analysis and Knowledge Extraction System (cTAKES).[61] Extracted mentions are further mapped to standard medical concepts that are represented by concept unique identifiers (CUIs) in the Unified Medical Language System (UMLS) [62] ontology.

Step 3. Topic modeling: We summarized mapped concepts to topics using Latent Dirichlet Allocation (LDA).[63] LDA is a probabilistic generative model for topic modeling. It represents each document as a mixture over latent topics, where each topic is modeled as a distribution over words.

Step 3a. Mention replacement: We replaced each extracted mention in the posts with its mapped CUIs and discarded all the other words in the posts. We will discuss this step in more detail in section 'Topic modeling' later.

Step 3b. Topic modeling using LDA: Given the corpus of mapped CUIs, LDA generates documents-topics and topics-CUIs probability distributions. We will discuss this step in more detail in section 'Topic modeling' later.

Step 3c. Analysis and evaluation: We further analyzed these distributions to derive a list of topics using most representative mentions, and summarized extracted mentions for each dataset. We will discuss this step in more detail in section 'Results: LDA topics' later.

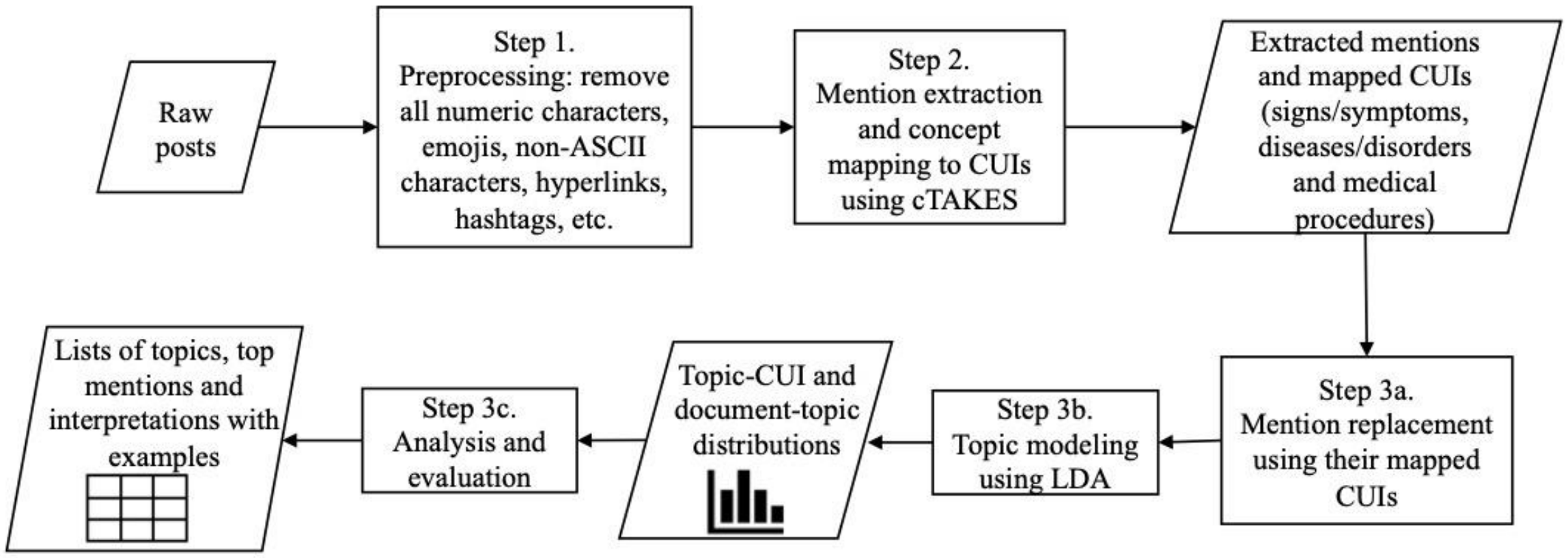

Figure 1: Pipeline for Breast Implant Illness Social Media Analysis

## *Data Preprocessing*

We used the NLTK tokenizer [64] to tokenize the raw texts for each dataset. Out of the obtained tokens, we removed the stop-words (most frequently occurring, function words such as conjunctions, prepositions, determiner, etc.) using the NLTK English stop-words list. Since stop-words carry little or no information on our topics of interest in BII, they can be safely removed as typically done in NLP. We also removed all the numeric characters, emojis, non-ASCII characters, hyperlinks, hashtags and Instagram handles using regular expression matching, and converted all the remaining tokens into lower case to unify different cases for downstream processing.

## *Mention extraction and concept mapping*

Mention extraction refers to the extraction of words/phrases that convey a medical concept. We used the cTAKES tool for mention extraction. The cTAKES tool is an open-source NLP tool for clinical information extraction from unstructured clinical texts. cTAKES extracts mentions (i.e., words/phrases that convey a medical concept) from posts and maps those mentions to standard medical concepts. In doing so, it also categorizes each extracted mention into one of five cTAKES categories: sign/symptom, disease/disorder, medication, procedure and anatomy, that is, while cTAKES is extracting the mentions, it also automatically classifies the mentions into one of the five categories. For example, in the sentence “Over the years my tinnitus, has become worse to almost debilitating levels”, cTAKES extracts “tinnitus” as a mention of sign/symptom category. Below, we discussed how to configure cTAKES in detail.

We used the fast-dictionary-lookup annotator in cTAKES to extract mentions from the processed data. This annotator identifies and extracts mentions in texts and normalizes them into CUIs in the UMLS standard medical ontology. This normalization of extracted mentions into CUIs is referred to as concept mapping. Each CUI in the UMLS ontology uniquely identifies a medical concept. Hence, we represented extracted mentions using standard medical concepts of those CUIs that cTAKES maps the mentions to. We configured the annotator to use exact string match and to use all-term-persistence property. Thus, the annotator retains all terms irrespective of the semantic properties of each term. For example, for the text “back pain”, the annotator annotates the generic term “pain” as well as the precise term “back pain”. We chose to use the all-term-persistence property to retain maximum information with respect to precise and generic medical concepts. Finally, the annotator stores the generated annotations in XMI files.

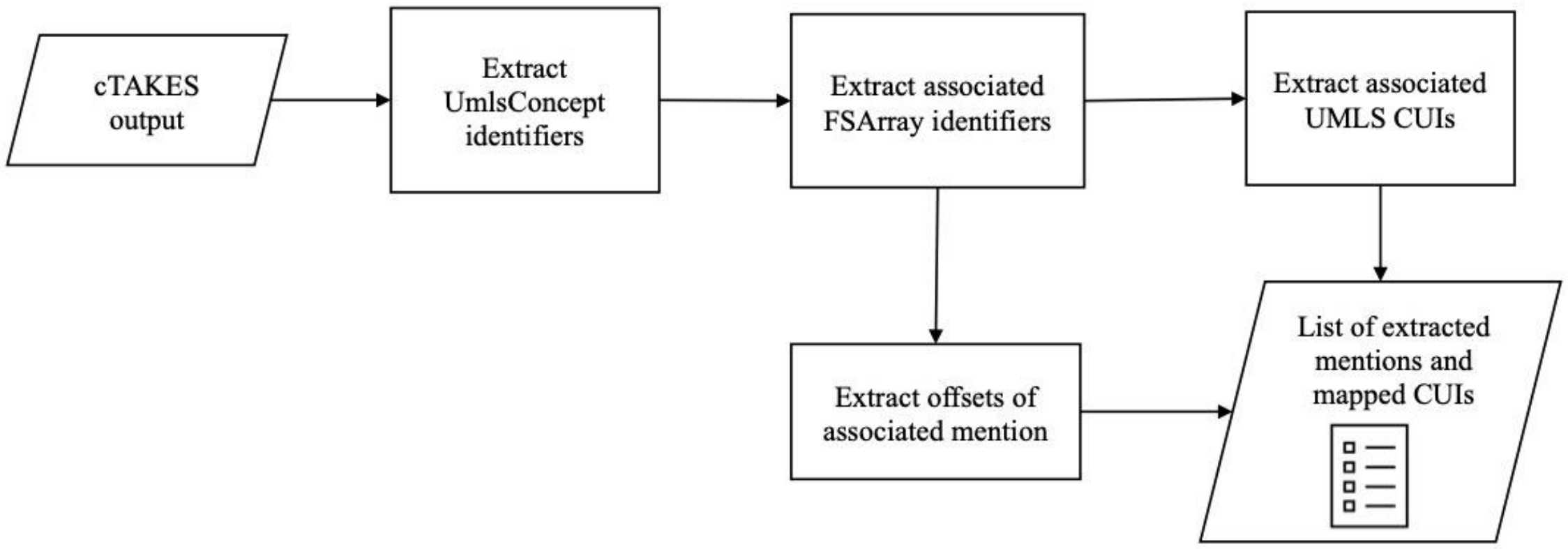

Figure 2: Pipeline for obtaining annotations out of cTAKES

In order to obtain the annotations in a human-readable format from the XMI files, we performed the following steps as shown in Figure 2. We used a custom interpreter to process XMI files produced by cTAKES and to obtain mappings between mentions and CUIs out of cTAKES. We first looked for *UmlsConcept* XML identifiers in the XMI files, where each *UmlsConcept* XML identifier is generally grouped under the *FSArray*, and each *FSArray* is associated with a single ontology concept and the category of the concept. Each such concept is assigned one category out of five cTAKES categories: sign/symptom, disease/disorder, medication, procedure and anatomy. Each ontology concept is further associated to a UMLS CUI and a *ontologyConceptArr* identifier. It must be noted that a mention can be mapped to multiple CUIs. For example, the mention "allergic reaction" is categorized as sign/symptom but mapped to two different CUIs "C1527304" and "C0020517". Then, we extracted those ontology concepts that describe any of these categories: diseases/disorders, signs/symptoms and medical procedures. Finally, we used the *begin* and *end* markers associated with each *ontologyConceptArr* identifier to obtain the position of the annotated mention in the input post. In this work, we are only interested in the first three categories (i.e., sign/symptom, disease/disorder and procedure) in order to understand BII-related issues. Hence, we only used the mentions that are categorized into either of these three categories.

## *Topic modeling*

In order to conduct topic modeling, we processed the posts as follows: we substituted each mention in the posts with its mapped CUIs and discarded all the other words in the posts, which were considered as non-medical concepts by cTAKES or not among the three categories of our interest. If a mention was mapped to multiple CUIs, we replaced that mention with the multiple CUIs. If multiple mentions were mapped to a same CUI, we replace all such mentions with the CUI. In this way, each post was represented as a bag-of-CUIs instead of a collection of mentions as the input to the topic modeling, and our vocabulary consisted of CUIs. Upon topic modeling, we interpreted the topic-CUI distribution to derive the topics.

We used Latent Dirichlet Allocation (LDA) [63] to learn the topic distributions of each post and the CUI distributions of each topic. LDA is a generative probabilistic model for modeling topics within a document corpus. LDA models each document in the corpus as a mixture of latent topics, where each topic is modeled as a distribution over words in all the documents. LDA derives the optimal distributions via maximizing the likelihood of observing the corpus following the perspective distributions. A brief description on LDA is provided in the Supplementary

Materials. In our experiments, a bag-of-CUIs generated as above was used as a document in LDA, and the CUIs were words in the document. We used the lda-c software [65], a very efficient implement of LDA method, to conduct the topic modeling.

When LDA is used for topic modeling for general documents (e.g., news, scientific literature), words and their frequencies in the documents are used in LDA. However, in our analysis, we aim to understand the medical concepts related to BII from social media texts. Different words may indicate a same medical concept. For example, joint aches, painful joints, arthralgia, aching joints all indicate joint pain and are associated with a single medical concept represented by a single CUI. Therefore, instead of using words, we used the medical concepts, which are presented by CUIs, in our LDA analysis. Since multiple words indicating a same medical concept can be mapped to a same CUI, using CUIs can also aggregate and strengthen the information from the multiple words, compared to using words, which may be sparse and thus not easy to learn topics from.

## Results

### cTAKES annotations

Table 2: Statistical Summary of Annotations out of cTAKES

| Dataset | #cwords[a] | #annots[b] | #maps[c] | #M[d] | #C[e] | #M/#C[f] | #C/#M[g] | #S[h] | #D[i] | #P[j] |
|---|---|---|---|---|---|---|---|---|---|---|
| BIIweb | 24,034 | 2,186 | 661 | 640 | 475 | 1.39 | 1.03 | 385 | 149 | 106 |
| HealingBII | 163,352 | 11,080 | 1,740 | 1,685 | 1,177 | 1.48 | 1.03 | 891 | 503 | 292 |
| IG-BII | 3,116,966 | 185,339 | 5,694 | 5,530 | 2,871 | 1.98 | 1.03 | 3,049 | 1,549 | 932 |

In this table, [a]#cwords indicates the total number of words recognized by cTAKES.
[b]#annots indicates the total number of extracted mentions belonging to the three semantic types (i.e., signs/symptoms, diseases/disorders and medical procedures).
[c]#maps indicates the number of unique mention-CUI mappings.
[d]#M indicates the number of unique extracted mentions.
[e]#C indicates the number of unique mapped CUIs.
[f]#M/#C indicates the average number of extracted mentions mapped to a given CUI.
[g]#C/#M indicates the average number of CUIs mapped to an extracted mention.
[h]#S indicates the number of unique extracted mentions that are mapped to the signs/symptoms category.
[i]#D indicates the number of unique extracted mentions that are mapped to the diseases/disorders category.
[j]#P indicates the number of unique extracted mentions that are mapped to the medical procedures category.

Table 2 presents the summary statistics on the annotated mentions and their mapped CUIs by cTAKES. In BIIweb, cTAKES extracted 2,186 mentions and mapped them to 475 unique CUIs. In HealingBII, cTAKES extracted 11,080 mentions and mapped them to 1,177 unique CUIs. In the largest dataset IG-BII, cTAKES extracted 5,530 unique mentions and mapped to 2,871 unique CUIs. Note that a same mention can be mapped to multiple CUIs and can have multiple categories (each CUI has only one category). For example, the mention “flashes” is mapped to two different CUIs and then two different categories: diseases and medical procedures. Table 2 also presents the statistics of each category of extracted mentions. For each dataset, most of the extracted mentions are categorized as signs/symptoms by cTAKES.

In order to determine if cTAKES can sufficiently extract relevant mentions, we performed a manual annotation and compared the two lists of extracted mentions: one out of cTAKES and the other out of the manual annotation. We randomly sampled 50 posts from each of the three datasets, and manually annotated those posts. Upon manual annotation, we extracted mentions (words/phrases) that convey concerns and experiences of social media users involving BII-related symptoms, diseases and medical procedures. For a random sample of 50 posts ($l_{avg}$ = 134.18) from BIIweb, we obtained a total of 575 mentions out of the manual annotation, and 637 mentions out of cTAKES. Out of these mentions, there are 479 in common. Note that each mention is associated with a post identifier and a character offset. A mention is considered to belong to both the lists if the same mention occurs in both lists with the same post identifier and character offset. We found that 83.30% (479 out of 575) of manually annotated mentions are covered by cTAKES. This high coverage demonstrates that cTAKES can capture most of the relevant medical concepts. On the other hand, 75.20% (479 out of 637) of annotated mentions by cTAKES are covered by the manual annotation. This further demonstrates that most of annotated mentions out of cTAKES can be confirmed by the manual annotation. Similarly, for a random sample of 50 posts ($l_{avg}$ = 80.02) from HealingBII, 69.53% (194 out of 279) of manually annotated mentions are covered by cTAKES; 70.39% (194 out of 276) of mentions annotated by cTAKES are confirmed by the manual annotation. For a random sample of 50 posts ($l_{avg}$ = 121.00) from IG-BII, the corresponding values are 75.21% (182 out of 242) and 70.39% (182 out of 283), respectively. According to the high overlap in the results between manual annotation and cTAKES across multiple datasets used in our study, it is reasonable to assume that cTAKES is a decent surrogate of manual annotation for BII study through social media data.

## LDA topics

In order to identify the best topic models, we used grid search to identify the best parameter values for the Dirichlet prior $\alpha \in \{0.01, 0.05, 0.1, 0.5, 1, 1.5, 2, 5, 10, 15, 20, 25\}$ and the number of topics $K \in \{3, 4, 5, 10, 15, 20\}$. In order to evaluate topic models, we analyzed each LDA topic modeling result for every combination of $\alpha$ and $K$ values corresponding to low perplexity scores.[63,66,67]

For each topic modeling result, we analyzed the document-topic and topic-CUI probability distributions to derive topics and their respective top-10 representative mentions. The top-10 representative mentions for a given topic are the most frequent mentions corresponding to the top-10 CUIs with the highest probabilities of belonging to the topic. Note that multiple mentions can be mapped to a given CUI (Table 2). We only presented the most frequent mention because all the mentions mapped to a same CUI have similar semantics. We further evaluated the quality of a topic modeling based on how well the derived topics summarize the most representative mentions. We analyzed each LDA topic modeling result for every combination of $\alpha$ and $K$; and chose the one where the derived topics were distinct and best summarized the most representative mentions. Finally, we identified distinct and meaningful topics using (i) $K = 4$ and $\alpha = 10$ for BIIweb, (ii) $K = 5$ and $\alpha = 10$ for HealingBII and (iii) $K = 5$ and $\alpha = 1.5$ for IG-BII. We observed that with higher $K$ values, the most representative mentions were similar across topics. Hence, the derived topics were not distinct and were difficult to interpret.

Table 3: Derived Topics in BIIweb

| topic | top-10 mentions | interpretation |
|---|---|---|
| 1 | testing (2.34); illness (4.46); problem (2.82); work (1.17); swollen (0.78); drains (0.61); feel common (2.51); fatigue (1.82); exhausted (0.39); sensitivity (0.95) | common signs/ symptoms |
| | Example: *"I had silicone implants done 5 years ago, three years ago after going to the doctor with extreme fatigue (I was sleeping 14-16 hours a day and was still exhausted)"* | |
| 2 | breast implant (6.8); removal (1.3); cancer (0.95); autoimmune (0.95); infection (0.87); scleroderma (0.39); pain (3.68); diagnosis (0.3); alcl (0.3); breast cancer (0.3); | diseases/ disorders |
| | Example: *"I had stage 4 breast cancer and had chemo and radiation. I tried to have my breast implants removed due to pain... Then I had an acute infection occur a month and a half after they put the new implants in and they were forced to perform an emergency removal of the newer implants. I have had all the symptoms of breast implant illness – even after their removal."* | |
| 3 | breast implant (6.8); illness (4.46); toxicity (1.17); foreign body (0.87); heal (0.78); support (0.65); rupture (0.52); cancer (0.95); awareness (0.35); inflammation (0.56) | toxicity |
| | Example: *"…I never had a problem until 2006 at which time I thought something had happened however, my surgeon said I must have just pulled a muscle and that the implants seemed fine. Now that surgeon is old and the shop is closed up. I have been suffering for the past 13 years with arthritis, fatigue, brain fog, inflammation, hormone imbalances, and adrenal fatigue…"* | |
| 4 | pain (3.68); feel (2.51); fatigue (1.82); back pain (0.87); illness (4.46); joint pain (0.56); worse (0.65); anxiety (0.52); ear ringing (0.39); headache (0.39) | pain and stress - related disorders |
| | Example: *"It wasn't until 2017 where I started to experience anxiety and panic attacks (which I didn't know I was having at the time). With that, along came crazy headaches, feeling dizzy, sick, lightheaded and my right eye would always be swollen and never knew why."* | |

Table 4: Derived Topics in HealingBII

| topic | top-10 mentions | interpretation |
|---|---|---|
| 1 | rupture (1.34); supported (0.87); read (1.17); suffering (0.87); happy (0.6); mastectomy (0.46); work (0.96); scare (0.77); reconstruction (0.41); mri (0.72); | surgeries and procedures |
| | Example: *"Double mastectomy in 2015. Reconstruction process with expanders then permanent 1000 ml saline implants in early 2016. After that was 9 procedures, a hysterectomy and now MANY health problems."* | |
| 2 | pain (3.91); joint pain (0.79); fatigued (0.96); ailment (4.7); removal (0.84); hair loss (0.52); headache (0.47); muscle ache (0.34); rash (0.39); infection (0.84) | pain and other signs |
| | Example: *"In addition to the neuromuscular spasms and pain, I've suffered with incapacitating chronic fatigue, BRAIN FOG and confusion (yes, even while driving), loss of vision and hearing, vertigo, mysterious skin rashes, hair loss, migraines, ..."* | |
| 3 | problem (2.64); cancer (0.9); autoimmune (0.57); breast cancer (0.38); scars (0.35); treatment (0.43); diagnose (0.29); autoimmune disorder (0.27); lupus (0.29); arthritis (0.26) | cancer and other disorders |
| | Example: *"I had capsules form on both breasts from about 2010. I got sick with BII symptoms from 2005 with lots of infections required intravenous and oral antibiotics. My environmental and drug allergies got worse, onset of arthritis, skin rashes, autoimmune symptoms, started growing low grade cancers,..."* | |
| 4 | breast implant (3.85); ailment (4.7); toxicity (3.05); healing (1.56); capsulectomy (0.64); infection (0.84); inflammation (0.39); detoxification (0.32); foreign object (0.25); bleed (0.23) | toxicity |
| | Example: *"Some women with silicone toxicity have bruising and bleeding problems. If I was you I would try and have the lymph node localized and checked for silicone and removed if it is contaminated beyond detoxing much like a silicone granuloma is removed."* | |
| 5 | emotion (3.7); think (2.26); feel (0.84); normal (0.65); anxiety (0.5); ill (0.61); sensation (0.33); tired (0.28); sores (0.27); depression (0.33) | mental health |
| | Example: *"Even more heartbreaking and discouraging, has been the emotional pain of not being able to freely play with her on the floor due to hip and knee pain, along with leg and foot spasms… but I struggle with many feelings of failure as a wife and mother due to physical limitations."* | |

Tables 3, 4 and 5 present the top-10 representative mentions, the frequencies of CUIs corresponding to the mentions (in %), and the interpretations of the topics indicated by the mentions (e.g., common signs/symptoms). Note that the frequencies of CUIs are among all the posts, not only in those posts with the highest probability belonging to a certain topic. We presented these frequencies because each post has a certain probability of belonging to a certain topic, and thus frequencies among all the posts should better represent the topic information across all the posts. These tables also present the examples of posts that have high probabilities of belonging to the respective topic. In the examples, the mentions that have high probabilities of belonging to the corresponding topics are underlined. Note that we used CUIs in LDA to derive topic and word distributions (as discussed in the section 'Methods – Topic modeling'), but we present the most frequent mentions (with clear semantics) that were mapped to respective CUIs (which are identifiers without semantics) in these tables. The mentions in these tables are sorted based on the probabilities of their corresponding CUIs belonging to the respective topics. Please note that these probabilities are not presented in the tables (they are not the frequencies presented in the tables). Therefore, each topic is represented by its most representative mentions and thus summarizes such mentions. For example, we interpret a topic as pain and other signs if there are significant number of mentions related to pain such as neck pain, chest pain, headache, etc. Please note that the topics are not sorted and the first columns in Tables 3—5 are only nominal identifiers. Below, we discussed the derived topics out of LDA for BIIweb and HealingBII datasets from the original posts. Note that two topics can still share a same representative mention with different probabilities in LDA.

Table 3 presents the topics in dataset BIIweb. Although BIIweb is the smallest dataset (Table 1), we were still able to identify four distinct topics with the most representative mentions such as fatigue, infection, toxicity and anxiety. Table 4 presents the topics in dataset HealingBII. HealingBII shares some common topics/representative mentions as those in BIIweb. For example, pains, cancers and toxicity are common across these two datasets. However, a focused topic unique in HealingBII is on surgeries and procedures, where people (mostly patients) discuss the procedures among themselves and share related experiences. Another unique topic in HealingBII is on mental health.

In addition to physical symptoms, individuals reported significant emotional and mental difficulties such as depression and expressed their serious symptoms in social media. Table 5 presents the topics in dataset IG-BII. IG-BII is the largest dataset (Table 1) and has significantly more posts than the other two. We observed that cancers, mental health and toxicity emerged as significant topics in this large dataset, quite consistently with those in HealingBII. In IG-BII, people also discussed their recovery process from issues/events associated with breast implant illness. We identified from these three datasets frequent mentions of rupture, pains and fatigue. We also identified mentions of cancer, lupus and autoimmune disorders. Please note that Table 3 has four topics for BIIweb, but Table 4 and 5 have 5 topics for HealingBII and IG-BII, respectively. This is because the number of topics is determined by how distinct the topics are, not by a pre-specified number of topics.

Table 5: Derived Topics in IG-BII

| topic | top-10 mentions | interpretation |
|---|---|---|
| 1 | heal (1.46); working (0.9); weighted (1.05); able (0.99); rest (0.37); stress (0.29); exercise (0.28); therapeutic (0.35); sleep (0.36); run (0.23) | physical health |
| | Example: *"It's been 14 months since my explant. The journey to healing hasn't been an easy one due to setbacks and relapses but better than daily anaphylaxis from getting cold, food, smells, crying, exercise and stress, then add angina attacks from anaphylaxis."* | |
| 2 | malignancy (1.1); removal (0.96); scar (0.75); capsulectomy (0.68); rupture (0.43); ciactrice (0.43); alcl (0.41); augmentation (0.37); lymphoma (0.35); removal of implants (0.29) | cancer and medical procedures |
| | Example: *"The ugly side of breast implants. It's not a matter of IF you will get sick.... it's WHEN. implants leak toxic heavy metals without rupture It's called a gel bleed. Women with implants are 3 times more likely to develop brain, lung and lymphatic cancer than women with implants."* | |
| 3 | loving (2.43); happiness (2.11); emotion (1.64); think (1.05); feel (0.87); scare (0.55); confidence (0.35); tired (0.38); emotional (0.27); sensation (0.33) | mental health |
| | Example: *"I was scared of looking incomplete. After much deep, inner work on myself, I realized that my worth wasn't dependent on what I looked like or how big my chest was. I realized that true happiness would come from 100% acceptance of what and who I was,"* | |
| 4 | breast implant (7.21); ailment (5.67); toxicity (1.67); aware (0.96); felt worse (0.36); test (0.64); foreign body (0.45); alone (0.33); suffering (0.21); complication (0.2) | toxicity |
| | Example: *"...We get toxic from the chemical makeup of the silicone, the toxic chemicals that are released when the shell degrades, sick from rupture and sometimes mold."* | |
| 5 | pain (2.52); inflammatory reaction (0.89); fatigue (0.83); anxiousness (0.72); allergy (0.43); depression (0.37); joint pain (0.33); autoimmune disorder (0.32); swell (0.43); infection (0.31) | common disorders |
| | Example: *"For three years doctors have been unable to diagnose or explain my upper body weakness, hand pain and general inflammation. I have suffered with periods of high inflammation, debilitating fatigue, migraines, unable to lose weight, insomnia, low libido, body and joint pain, hair loss, dry skin, dry eyes, brain fog, etc."* | |

Table 6 presents the top-10 representative mentions, the frequencies of CUIs corresponding to the mentions (in %), and the interpretations of the topics on the unified dataset – combining all the three datasets BIIweb, HealingBII and IG-BII. We obtained the unified dataset by combining all the posts from the three datasets into one corpus. In order to perform topic modeling, we processed the posts in the unified dataset in the same way as we processed the posts in the individual datasets (discussed in the section 'Methods – Topic modeling'). Upon topic modeling, we identified five distinct topics using $K = 5$ and $\alpha = 1.5$. We observed that physical health, cancers, mental health, toxicity and common disorders emerged as significant topics in the unified dataset, quite consistently with those in IG-BII. This is because IG-BII is the largest dataset out of the three and comprise more than 90% of the unified dataset. We also identified common concerns such as pain, allergy, depression, weight gain, cancer, inflammation and toxicity issues from the individual and the unified datasets. This implies that the above mentions are frequently associated with breast implant illness.

Table 6: Derived Topics in the Unified Dataset

| topic | top-10 mentions | interpretation |
|---|---|---|
| 1 | working (1.45); ate (0.92); weight (0.79); runs (0.40); thinking (2.68); exercise (0.25); talk (0.50); walking (0.35); nutrition (0.15); move (0.28); | physical health |
| | Example: *"...I'm now healthier than I have been in the last 7 years of my life!... I explanted in Feb of 2018, a few months after explant, I gained my weight back and found a love for true self care and working out."* | |
| 2 | illnesses (4.45); cancer (0.87); ruptures (0.77); removal (0.76); awareness (0.73); suffers (0.83); capsulectomy (0.54); autoimmune (0.52); breast augmentation (0.30); augmentation (0.28); | cancer and medical procedures |
| | Example: *"I was diagnosed with breast cancer at the young age of 30 and ended up with a double mastectomy as part of that process... now 10 years later I have just 15 weeks ago had my implants removed. They had ruptured, were toxic and giving me health issues"* | |
| 3 | feel (5.94); loved (2.97); thinking (2.68); happier (1.64); feelings (1.47); afraid (0.66); confidence (0.27); support (0.79); able (0.77); alive (0.17); | mental health |
| | Example: *"When I found out I was sick and I had to tear apart my body to get better I never thought I'd be happy with myself again. I am 4 weeks post op and feeling more happy and healthy than ever. I was worried I'd never be loved again."* | |
| 4 | heal (2.26); scars (0.58); scarred (0.33); drain (0.26); toxic (1.97); sights (1.25); inflammation (0.68); bulge (0.36); tenderness (0.20); red (0.15); damage (0.16); | common signs/symptoms and toxicity |
| | Example: *"I was so worried about how red and raised up my scars were... then they got really inflamed, sore and raised up around 3 weeks and i was really stressed over it. then overnight the inflammation and redness went down..."* | |
| 5 | pain (2.09); tired all the time (0.69); anxiety (0.57); joint pain (0.46); alopecia (0.39); weight gain (0.37); allergies (0.35); depression (0.29); pain back (0.23); head ache (0.22) | common disorders |
| | Example: *"Before I had the explant I had many unexplained symptoms (brain fog, joint pain, back and neck pain, tired all the time, psoriasis, afib, just to mention a few) since I awoke from surgery I have had absolutely no neck, back or joint pain."* | |

Table 7: Distributions of Posts over Topics

| dataset | topic | %posts |
|---|---|---|
| BIIweb | common signs or symptoms | 33.15 |
| | diseases or disorders | 14.97 |
| | toxicity | 26.74 |
| | pain and stress-related disorders | 25.14 |
| HealingBII | surgeries and procedures | 37.13 |
| | pain and other signs | 11.51 |
| | cancer and other disorders | 11.51 |
| | toxicity | 26.31 |
| | mental health | 13.54 |
| IG-BII | physical health | 38.98 |
| | cancer and medical procedures | 13.42 |
| | mental health | 16.83 |
| | toxicity | 18.68 |
| | common disorders | 12.09 |
| unified | physical health | 15.31 |
| | cancer and medical procedures | 34.21 |
| | mental health | 25.58 |
| | common signs/symptoms and toxicity | 12.96 |
| | common disorders | 11.94 |

Table 7 presents the percentage of posts per topic, where a post $\boldsymbol{d}$ is considered as belonging to a topic $z$ if among all topics that $\boldsymbol{d}$ has, $z$ has the highest probability. Although the distributions are not completely consistent across datasets, toxicity remains a notable topic among all the datasets. This indicates the common issues significantly associated with breast implant illness. Also, pains, cancers, mental health and other disorders are substantially associated with breast implants.

## Discussion

In order to understand signs, symptoms and diseases/disorders associated with breast implant illness, a condition reported primarily in social media rather than medical reports, we collected social media posts and analyzed them using NLP and topic modeling. We extracted mentions related to signs/symptoms, diseases/disorders and medical procedures using cTAKES, mapped them to standard medical concepts, and summarized the mapped concepts to topics using LDA. We found that mentions such as rupture, infection, inflammation, pains and fatigue were common self-reported issues. We also found that mental health related concerns such as stress, anxiety and depression as well as diseases like cancers and autoimmune disorders were common concerns. Note that cTAKES is able to extract medication and anatomy information as well, but they were not used in our LDA analysis given that the objective of our study is not to study medications used or anatomy related to BII.

In our method, we relied on cTAKES and the rich UMLS dictionary to extract all relevant mentions including their lexical variants (synonyms, abbreviations, paraphrases). In order to determine if cTAKES can sufficiently extract relevant mentions, we performed a manual annotation to extract all relevant mentions and compared those with extracted mentions out of cTAKES. We found that cTAKES can sufficiently capture relevant medical concepts, quite comparable to the manual annotation. It is worth noting that we did not evaluate the performance of our mention extraction module on all the posts of each dataset, which is typically done using precision and recall metrics when there are ground-truth labels associated with each mention. However, in order to have such labels, it requires careful manual annotations based on domain knowledge on breast implant illness. Unfortunately, such domain knowledge on complications, symptoms and other issues associated with/caused by breast implant illness is not fully available. Actually, our goal in this study is to provide useful information from social media data that could help complement what we currently know. Therefore, in this preliminary study, we use all the annotated mentions, assuming that cTAKES enables high-quality annotations.

We acknowledge that cTAKES might not be able to extract all relevant mentions from our social media datasets. This is because cTAKES was originally designed for extraction of medical entities from clinical notes, which have very different wording and writing styles compared to social media data. As social media data comprise informal phrases, short ambiguous texts, emoticons and a wide range of lexical variants corresponding to a single concept, cTAKES may not work flawlessly on social media data, although we observed reasonable output out of cTAKES. We also observed that cTAKES often associates a single mention with multiple CUIs belonging to the same category. We think it is due to the presence of multiple mappings for a given mention in the UMLS meta-thesaurus. Regardless, the extracted mentions and the mapping of mentions to UMLS CUIs as generated by cTAKES are used for topic modeling without any manual verification or evaluation. In the future research, we will develop a detailed guideline to further evaluate extracted mentions before using them in topic modeling.

Our study also has some limitations. First, LDA is an unsupervised learning technique, in which the number of topics ($K$) is assumed to be known a-priori. However, it is difficult to accurately estimate $K$ for a given dataset. In our study, we used grid search to try different $K$ values. Even though, without full domain knowledge, it still remains non-trivial to evaluate the LDA results for each $K$ value. In our study, we selected the topics based on $\alpha$ and $K$ values. We did not use perplexity [63,66,67], a widely used metric in topic modeling to select topics, because as studied in literature (e.g., Chang *et al*. [68]), perplexity often does not correlate well with topic interpretability, and in our case, the lowest perplexity does not always enable intuitive or meaningful topics. In the future research, we will develop more rigorous ways to select the number of topics and to evaluate topic modeling results. In our current study, we did not do a sentiment analysis on the posts in order to understand the positive or negative opinions expressed in the posts. We plan to include this process before topic modeling so as to generate cleaner dataset for topic modeling.

It is always worth noting that social media data could be of variable quality (e.g., misspelling, misconception, biased opinions), particularly compared to medical literature data. Anyone can post to social media, and so the derived content may be from individuals who may have other implant specific issues such as capsular contracture or implant infection. Thus, understanding diseases, disorders, symptoms, signs, etc., associated with a drug, disease or medical procedure from social media data always runs into risks of confounders or errors. However, given that the medical knowledge and literature on breast implant illness have not been well established, and the related concepts are not well defined or well accepted, using social media data to understand emerging issues could be a meaningful starting point. Still, any findings from social media data require rigorous evaluation and validation based on medical and biological knowledge, experiments and clinical practice, etc. In addition, we only analyzed three, though the most relevant and prolific, websites dedicated to BII discussions. Additional, more comprehensive analysis on social media data of a much larger scale would be beneficial to better understand BII from a larger, diverse population. Sentiment analysis over social media data could be another valuable analysis to enable more insights on health experience of users/patients and their emotions/feelings. We will consider sentiment analysis in our future research when BII is better understood and we can accurately annotate social media data.

This study has important implications for future methodological work and clinical research. Future methodological research on NLP could include causality inference between breast implant illness and symptom/sign mentions from social media to understand their relations, etc.

Our findings could provide the relevant domains for clinical research studies that are seeking to develop measures of BII, and to identify its causes. More specifically too, our results provide a patient derived definition of BII which can be useful to clinicians treating patients with BII concerns in order to use this patient-centered language. Our methods and informatics strategies applied in this study would also provide working examples for analyzing other emerging but not well-defined illnesses from social media data.

Our analysis over social media data identifies mentions such as rupture, infection, inflammation, pains and fatigue that are common self-reported issues on social media sites dedicated to BII. In addition, our analysis shows that a significant number of the user comments and posts are also concerned with mental and physical health, and toxicity issues after breast implants. The findings from our study could be used to further the scientific study of BII as well as the care of patients presenting with the described symptoms by allowing clinicians to develop a patient-centered language to better approach patients with concerns. Our study provides the first analysis and derived knowledge of BII from social media using NLP techniques, and demonstrates the potential of using social media information to better understand emerging illnesses.

## Acknowledgements

Xia Ning conceived the research, obtained funding for the research, and supervised Vishal Dey; Peter Krasniak, Minh Nguyen and Clara Lee provided substantial medical background and insights; Vishal Dey and Xia Ning conducted the research, including data curation, methodology design and implementation, analysis; Vishal Dey drafted the original manuscript; Vishal Dey and Xia Ning conducted the manuscript editing; Peter Krasniak, Minh Nguyen and Clara Lee reviewed the manuscript and provided constructive suggestions and feedbacks.

## Conflicts of Interest

The authors claim no conflict of interests.

## Abbreviations

BII: Breast Implant Illness
NLP: Natural Language Processing
cTAKES: Clinical Text Analysis and Knowledge Extraction System
CUI: concept unique identifier
UMLS: Unified Medical Language System
LDA: Latent Dirichlet Allocation

# Supplementary Materials

# A Pipeline to Understand Emerging Illness via Social Media Data Analysis: A Case Study on Breast Implant Illness

## Brief Description of LDA:

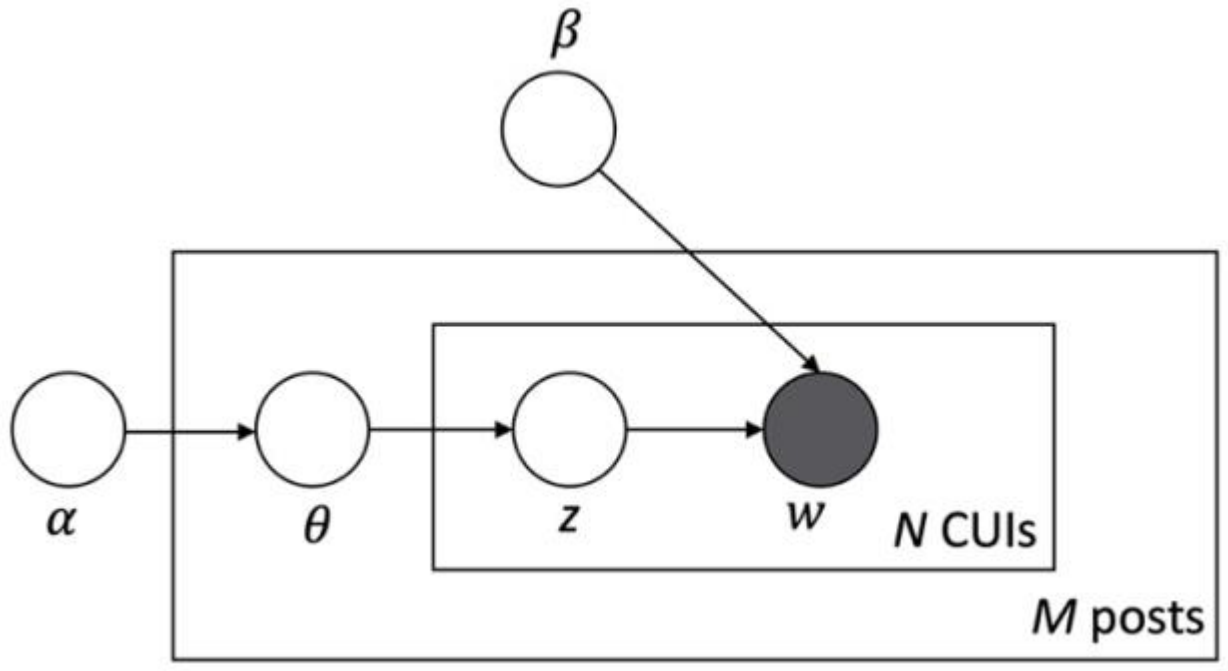

Figure S1: Graphical model of LDA

Here, we briefly describe Latent Dirichlet Allocation (LDA).[1] LDA is a generative probabilistic model that discovers latent topics in a document corpus. LDA assumes that a document of $N$ words $\boldsymbol{w} = \{w_1, w_2, \cdots, w_N\}$ is generated as follows: 1) a per-document distribution over topics $\boldsymbol{\theta} \in \mathbb{R}^K$ is first generated from a Dirichlet distribution Dirichlet($\boldsymbol{\alpha}$), where $\boldsymbol{\alpha} \in \mathbb{R}^K$ is the Dirichlet prior $\alpha_k \geq 0$ $(k = 1, \cdots, K)$ and $K$ is the given number of topics; 2) for each word $w_i$ in the document, a topic $z_i$ is generated from a multinomial distribution Mult($\boldsymbol{\theta}$); 3) a word distribution $\boldsymbol{\varphi}_i \in \mathbb{R}^L$ over topic $z_i$ is generated from a Dirichlet distribution Dirichlet($\boldsymbol{\beta}$), where $\boldsymbol{\beta} \in \mathbb{R}^L$ is the Dirichlet prior, $\beta_l \geq 0$ $(l = 1, \cdots, L)$ and $L$ is the number of words in the vocabulary; 4) given $\boldsymbol{\varphi}_i$, word $w_i$ is generated from a multinomial distribution Multi($\boldsymbol{\varphi}_i$). LDA assumes all the words $w_i$ in a document are independent given their $\boldsymbol{\varphi}_i$, and all the documents in the corpus are independent. Estimation on $\boldsymbol{\theta}$ and $\boldsymbol{\varphi}$ via maximum likelihood methods will enable document topics and the most probable words over the topics.